\documentclass[aps,prd,onecolumn,superscriptaddress]{revtex4}
\usepackage{graphicx}
\usepackage{amsmath}
\usepackage{amsmath,amsfonts,amssymb}
\usepackage{pdfpages}
\usepackage{graphics}
\usepackage{epsfig}
\usepackage{amsmath,amsfonts,amssymb}
\usepackage{braket}
\usepackage{mathbbol}
\usepackage[T1]{fontenc}
\usepackage{bbold}
\usepackage{amsmath}
\usepackage{amsmath,amsfonts,amssymb}
\usepackage{times}
\usepackage{braket}
\usepackage[T1]{fontenc}
\usepackage{bbold}
\usepackage{mathbbol}
\usepackage{graphicx}
\usepackage{times}
\DeclareGraphicsExtensions{.png,.pdf}

\usepackage{color}

\begin{document}
\baselineskip=12pt
\def\black{\textcolor{black}}
\def\red{\textcolor{black}}
\def\blue{\textcolor{blue}}
\def\green{\textcolor{black}}
\def\be{\begin{equation}}
\def\ee{\end{equation}}
\def\bea{\begin{eqnarray}}
\def\eea{\end{eqnarray}}
\def\orc{\Omega_{r_c}}
\def\om{\Omega_{\text{m}}}
\def\E{{\rm e}}
\def\bearst{\begin{eqnarray*}}
\def\eearst{\end{eqnarray*}}
\def\peleven{\parbox{11cm}}
\def\peffec{\peight{\bearst\eearst}\hfill\peleven}
\def\pspace{\peight{\bearst\eearst}\hfill}
\def\ptwelve{\parbox{12cm}}
\def\peight{\parbox{8mm}}

\title{Late time sky as a probe of steps and oscillations in primordial Universe }

\author{Mohammad Ansari Fard}
\address{Department of Physics, Sharif University of
Technology, P.~O.~Box 11155-9161, Tehran, Iran}

\author{Shant Baghram}
\email{baghram-AT-sharif.edu}
\address{Department of Physics, Sharif University of
Technology, P.~O.~Box 11155-9161, Tehran, Iran}

\begin{abstract}
The standard model of cosmology with nearly Gaussian, isotropic, scale invariant and adiabatic initial conditions describes the cosmological observations well.
However, the study of any deviation from the mentioned conditions will open up a new horizon to the physics of early universe. In this work, we study the effect of the oscillatory and step-like features in potentials of inflationary models in late time large scale structure observations. Mainly we study the matter power spectrum, number density of the structures, dark matter halo bias and specifically CMB lensing. We show that the oscillatory models can introduce some degeneracy with late time effects on BAO scale. We also conclude that high frequency oscillatory models which are favored by Planck data do not have significant effect on the non linear structure formation. Finally we show that inflationary models with step functions which deviates from the standard model in small scales $l \leq 1 Mpc$ can be constrained by future experiments via CMB lensing. We propose the idea that CMB lensing is a bias independent observation which can be used as a small scale physics probe due to distribution of the lenses in low redshifts. Meantime this model can alter the prediction of the cosmological model for the number density of small structures and can be used as a probable explanation for galactic scale crisis of $\Lambda$CDM.

\end{abstract}

\maketitle

\section{INTRODUCTION}
The cosmic microwave background (CMB) radiation measurements \cite{Ade:2013zuv,Ade:2015xua} and the large scale structure (LSS) surveys \cite{Tegmark:2003ud} indicate that the standard model of cosmology known as $\Lambda$CDM with its two unknown components of cosmological constant and cold dark matter (CDM) describes the observations well. The standard model of cosmology also encapsulates in it the initial conditions of early time curvature perturbations, which are nearly Gaussian, scale invariant, isotropic and adiabatic. The conditions are predicated by inflationary models in their simplest manifestation via the venue of cosmological perturbation theory\cite{Guth:1980zm,Linde:1981mu,Mukhanov:1990me}.
Despite the great triumph of the standard model, the physics of accelerated expansion of the universe, dark matter and early universe is still unknown. The physics of early universe can be examined and probed if we find any deviation from the standard initial conditions.
Consequently the observational verification of this deviation has a great importance.\\
The CMB observations are the dominant arena which early universe models are examined \cite{Ade:2015lrj}. However there is a decade or so that a great attention is focused on LSS as an alternative way to constrain the early universe physics \cite{Dalal:2007cu,LoVerde:2007ri,Komatsu:2009kd,Hanson:2009kg,Verde:2010wp, Desjacques:2010nn, Baghram:2013lxa, Mirzatuny:2013nqa, Abolhasani:2013vaa, Baghram:2014nha, Hu:2014hra,Hassani:2015zat}.
The main motivation of this work is to study the effects of inflationary models, which deviates from scale invariant initial power spectrum, on large scale structure observations. This is important because the standard picture of the structure formation indicates that the primordial anisotropies seen in CMB temperature maps has been sourced by the gravitational potential anisotropy in last scattering surface. This potential anisotropies grow due to the gravitational instability and become the late time structures in low redshift universe.\\
In this work we study the effect of features in primordial power such as a sharp feature in potential or
a resonance feature which is periodic in time and are introduced in inflationary models such as
axion monodromy, brane inflation and also in multi-field inflationary models \cite{Martin:2000xs,Danielsson:2002kx,Bozza:2003pr, Silverstein:2008sg,Kobayashi:2012kc,McAllister:2008hb}.
These features despite to their model building and theoretical motivations can be presented
as standard clocks which can be used to parameterize the scale factor $a(t)$ and put constraints on the type of early universe models \cite{Chen:2014cwa}, where a pioneer works on this subject and the effect of these models on LSS is presented by  Chen et al. \cite{Chen:2016vvw} and Ballardini et al. \cite{Ballardini:2016hpi}. In this work we study the effect of these models on LSS observables beyond the matter power spectrum which is studied in \cite{Chen:2016vvw,Ballardini:2016hpi}.\\
We track the imprints of inflationary models in angular power spectrum of CMB anisotropies.
We also study the effects of these models in LSS observations such as power spectrum and correlation function of matter perturbations. The number density of the structures are also discussed in this work. The advantage of LSS observations is  that they produce 3D maps of the perturbations that have more information than the 2D map of CMB. Also we can study the perturbations in sub-CMB scales and pin down some deviations which are out of the reach of the CMB experiments.
However we should note that the data in late time is noisy, non-linear and biased.
In this direction we study the effect of the early universe models on dark matter halo bias. Then we conclude that the observations which trace the distribution of dark matter directly and are bias independent are prominent ones for further studies. CMB lensing is a specific example of these observations, which we discuss it in this work with extend \cite{Lewis:2006fu,Ade:2015zua}.\\
In this work we set the standard cosmological parameters by Planck\cite{Ade:2013zuv}.
This is done for simplicity of the study, {in order to track the primordial effect on LSS, we assume that the late time evolution of the Universe after CMB is not altered by these modifications and the only effect which plays the crucial role is imprinted in primordial power spectrum. Accordingly we set $\Omega_m=0.316$ (matter density parameter), $\Omega_{\Lambda}=0.684$ (cosmological constant density parameter),  $H_0=67.27$ km/s/Mpc (present value of the Hubble parameter) and  $\tau=0.079$ (optical depth)\cite{Ade:2013zuv}.
The structure of this work is: In Sec.(\ref{Sec-1}), we introduce the inflationary models with oscillations and steps which are considered as representative models introducing local and non-local features. In Sec. (\ref{Sec-2}) we examined the imprint of features on late time sky observations such as matter power spectrum, correlation function, halo number density, halo-matter bias and redshift space distortion. In Sec. (\ref{Sec-3}) we consider CMB lensing as bias independent observation. We calculate the effect of the inflationary models feature on CMB lensing angular power spectrum. Also we forecast for future PRISM \footnote{Polarized radiation imaging and spectroscopy mission} like CMB experiments \cite{Andre:2013nfa}. Finally in Sec. (\ref{Sec-Conc}) we conclude and we discuss the future prospects of the work. In appendix (A), we study the CMB temperature anisotropy and LSS power spectrum. In appendix (B), we study the physics of number density with more detail and finally  In appendix (C), we summaries the CMB lensing Fisher analysis.

\section{Inflationary models with steps and oscillations}
\label{Sec-1}
In this section, we introduce the inflationary models that we are going to probe their effect on LSS data. The physics of the scalar perturbation is imprinted in the two point correlation function of the gauge invariant curvature perturbation ${\cal{R}}$ or interchangeably in it's power spectrum, which is parameterized as
\be \label{eq:power}
\bar{\cal{P}}_{{\cal{R}}}(k)=\frac{k^3}{2\pi^2}|{\cal{R}}_k|^2=A_s (\frac{k}{k_*})^{n_s-1},
\ee
where $\bar{\cal{P}}_{{\cal{R}}}$ is dimensionless power spectrum of curvature perturbation in the standard case of Gaussian, isotropic, adiabatic and nearly scale  invariant initial conditions. In the definition of the power spectrum $A_s$ is the amplitude of the dimensionless power in the pivot wavenumber $k_*$ and $n_s$ is the spectral index which shows the dependence of the power to the wavenumber.
The CMB anisotropy observations put strict constraints on these parameters in the CMB scales, $n_s=0.9645 \pm 0.0049$ and $\ln( 10^{10}A_s)= 3.094 \pm 0.034$ \cite{Ade:2015xua}, where in this work we will use these best fits for the standard case study. The constraints on the amplitude and spectral index of the primordial power is translated to the constraints on early universe inflationary models \cite{Ade:2015lrj}.
However as mentioned in the introduction, it worths to test extensions of the primordial initial condition and check the robustness of the prediction of the standard curvature  power spectrum. We should also note that the full sky CMB observations spans the angular moment scales of $\ell \sim 2- 3000$, which means that in sub-CMB scales $k > 0.3 h Mpc^{-1}$ ($k$ is the comoving wavenumber), we could have some deviations from the standard primordial power spectrum. \\
In this work, we will study the models that have high energy physics motivations and are also phenomenologically rich ones. In some cases they can address the galactic scale challenges of standard model of cosmology (i.e. $\Lambda$CDM).
The first modification which we consider in this work is the extension which is defined with logarithmic oscillations as
\be \label{eq:power-log}
{\cal{P}}^{log}_{{\cal{R}}}(k)=\bar{\cal{P}}_{{\cal{R}}}(k)\left[1+A_{log}\cos \left(\omega_{log}\ln(\frac{k}{k_*})+\phi_{log}\right)\right],
\ee
where $\bar{\cal{P}}_{{\cal{R}}}$ is the standard power low curvature perturbation power spectrum. The amplitude, frequency and phase $A_{log}$, $\omega_{log}$ and $\phi_{log}$ respectively are free parameters of the power spectrum extension. This type of oscillatory powers appears in non-Bunch Davis initial condition \cite{Martin:2000xs,Danielsson:2002kx,Bozza:2003pr} or in axion-monodromy models \cite{Silverstein:2008sg,McAllister:2008hb}.
\begin{figure}
\centering
\includegraphics[width=0.45\textwidth]{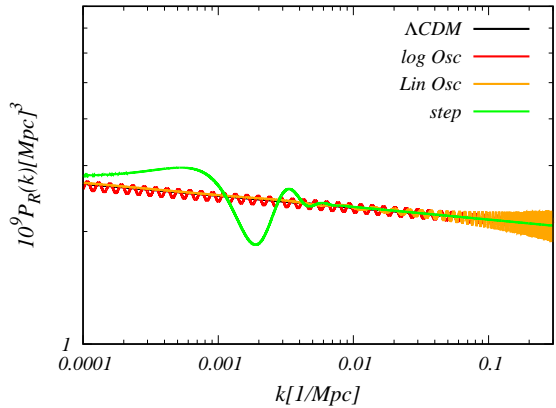}
\caption{Primordial power spectrum for $\Lambda$CDM (black), logarithmic (red) and linear (orange) oscillatory models
and also a step function model (green) is plotted versus the wavenumber.  The free parameters of the models are fixed by Planck data \cite{Ade:2015lrj}.}
\label{Fig:primordialpwer}
\end{figure}
In Fig.(\ref{Fig:primordialpwer}), the logarithmic oscillations primordial power spectrum is plotted (red line), which can be compared with the black solid line, which represents the $\Lambda$CDM standard initial condition power spectrum.
In this work, we study the matter power spectrum and the matter density variance of this model and we conclude that the logarithmic-oscillatory models have different and distinguishable features in late time observables.\\
The other type of the oscillatory modifications which appears in inflationary models is the linear oscillation with a scale dependence amplitude. These models can arise from boundary effective field theories \cite{Jackson:2013vka}. We use the parametrization of Meerburg, Spergel and Wandalt \cite{Meerburg:2013dla} as below
\be \label{eq:power-lin}
{\cal{P}}^{lin}_{{\cal{R}}}(k)= {\bar{\cal{P}}}_{{\cal{R}}}(k)\left[1+A_{lin}(\frac{k}{k_*})^{n_{lin}}\cos(\omega_{lin}(\frac{k}{k_*})^{1/p}+\phi_{lin})\right],
\ee
where $A_{lin}$, $n_{lin}$, $p$ and $\phi_{lin}$ are free parameters.
In the case, if we set $\omega_{lin}=p^2\Omega/(1-p)$, where $\Omega$ is a free parameter, we will recover the power spectrum introduced by Chen, Namjoo and Wang \cite{Chen:2014cwa} for primordial standard clock models.
In Fig.(\ref{Fig:primordialpwer}), we plot the linear oscillation power spectrum (in solid orange) in comparison with the standard case.\\
As a final example, we will study the inflationary model with a localized modification to the power spectrum.
Step-like features in the potential of an inflationary model, which is introduced in a specific small range of wavenumbers \cite{Adams:2001vc} or the change in the sound speed \cite{Achucarro:2010da,Nakashima:2010sa} will lead to this type of localized change in the scalar primordial power
spectrum. A general parameterization describing the step-like models was proposed by Miranda and Hu (2014)\cite{Miranda:2013wxa}, where we use the simplified manifestation of this model studied by \cite{Ade:2015lrj}
\be \label{eq:power-step}
{\cal{P}}^{step}_{{\cal{R}}}(k) = \exp\left[\ln {\bar{P}}_{\cal{R}}(k)+ I_0(k)+\ln(1+I_1^2(k))\right],
\ee
where $I_0$ and $I_1$ are defined as
\begin{eqnarray}
I_0(k) &=& A_s W_0(k/k_s) D(\frac{k/k_s}{x_s}), \\ \nonumber
I_1(k) &=& \frac{1}{\sqrt{2}}\left[\frac{\pi}{2}(1-n_s) + A_s W_1(k/k_{s})D(\frac{k/k_{s}}{x_s}) \right],
\end{eqnarray}
with the window functions
\begin{eqnarray}
W_0 &=& \frac{1}{2x^4}[(18x - 6x^3)\cos(2x) + (15x^2 - 9)\sin(2x)] , \\ \nonumber
W_1 &=& \frac{1}{x^4} \{3(x\cos x -\sin x) [3x\cos x + (2x^2-3)\sin x] \} ,
\end{eqnarray}
where $x=k/k_s$  and $D(x) = x /\sinh x$ is the damping function. We should note that the only parameter that we are interested in is the position of the step function, which is represented by $k_{s}$.
In Fig.(\ref{Fig:primordialpwer}), we plot the step function primordial power spectrum (in green solid line) in comparison with the standard case.
In Appendix(A), we will study the effect of these models on CMB angular power spectrum, where the strictest constraints come from. In the next section, we will study the effect of these models on late time observables.

\section{Late time Sky}
\label{Sec-2}
In order to study the effect of the deviation from standard initial conditions in LSS observations, we look at the perturbed universe in low redshifts and we will study the dynamics and statistics of perturbations. In the first subsection we study the interconnection of the matter and initial curvature power spectrum. In the second subsection, we will study the effect of inflationary models in number density of halos, in the third one we discuss the halo-matter bias and in the final subsection we study the redshift-space distortion effect and the correlation function of dark matter tracers.

\subsection{Primordial power spectrum vs matter power spectrum }
In order to study the evolution of the late time Universe we use the perturbed FRW metric in Newtonian gauge (we set $c=1$)
\begin{equation}
ds^2=a^2(\eta)\left[-(1+2\Phi(\vec{x},t))d\eta^2 + (1 - 2\Phi(\vec{x},t))d\chi^2\right],
\end{equation}
where $\Phi(\vec{x},t)$ is the Bardeen potential and the scalar degree of freedom.
We will look at the evolution of the Bardeen potential in late time Universe and its relation with primordial initial condition in this section.
The physics and statistics of the primordial curvature perturbation put its fingerprint on late time LSS via Bardeen potential.
The gravitational potential in Fourier space, $\Phi(k,z)$, is related to the primordial curvature perturbation ${\cal{R}}_k$ via an evolutionary equation which depends on redshift and wavenumber as
\be \label{eq:evo}
\Phi(k,z)=\frac{3}{5}T(k)D(z)(1+z){\cal{R}}_k,
\ee
where $T(k)$ is the transfer function, which imprints the evolution of the potential in horizon crossing and radiation-matter equality. The growth function $D(z)$ shows the redshift evolution and growth of the potential in late Universe. In standard $\Lambda$CDM model this growth function is scale independent. The normalization is defined in such a way that $D(z)(1+z)$ goes to unity in deep dark matter dominated era. In this work we use the standard fitting function of $\Lambda$CDM model for growth function. For the transfer function, we will use the function introduced by Eisenstein and Hu\cite{Eisenstein:1997ik}, which includes the baryonic effects on the matter power spectrum.
Now in order to see the connection of matter distribution function and density contrast with primordial curvature perturbation we use the Poisson equation in Fourier space
\be\label{eq:poisson}
k^2\Phi(k,z)=4\pi G\rho_ma^2\delta^*(k,z),
\ee
where $\delta^*=\delta+3{\cal{H}}(1+w)\theta_k/k^2$ is the gauge invariant density perturbation and $\theta_k=i\vec{k}.\vec{v}$ is the divergent of the peculiar velocity in Fourier space, ${\cal{H}}=aH$ is the conformal Hubble parameter and $\rho_m$ is the background matter density. In sub-horizon regime we can neglect the second term in definition of $\delta^*$ and replace the gauge invariant quantity with matter density contrast ($\delta^*\simeq\delta$). 
Using the evolutionary Eq.(\ref{eq:evo}) we can relate the matter density contrast to the primordial curvature perturbation
\be  \label{eq:deltam}
\delta_m(k,z)=\frac{2}{5}\frac{k^2 T(k)D(z)}{H_0^2\Omega_m}{\cal{R}}_k,
\ee
where $H_0$ is the present value of the Hubble constant. Accordingly we can find the dark matter power spectrum using Eq.(\ref{eq:deltam}) as
\be \label{eq:matterpower}
P_m(k,z)=\frac{4}{25}\frac{k^4T^2(k)D^2(z)}{H^4_0\Omega_m^2}P_{\cal{R}}(k),
\ee
where $P_{\cal{R}}(k)$ is the curvature power spectrum.
The matter power spectrum can be written in its conventional form as below
\be \label{eq:matterpower1}
P_m(k,z)=A_{lss}T^2(k)D^2(z)k^{n_s},
\ee
where $A_{lss}$ is the amplitude of matter power spectrum in late time for LSS, and $n_s$ is the spectral index of primordial perturbations. The dimensionless curvature perturbation power spectrum ${\cal{P}}_{\cal{R}}(k)$ is defined in Eq.(\ref{eq:power}). Accordingly the matter power spectrum amplitude $A_{lss}$ is related to the primordial power due to Eq.(\ref{eq:evo}) as
\be
A_{lss}=\frac{8\pi^2}{25}\frac{k_*^{1-n_s}}{H_0^4\Omega_m^2}A_s.
\ee
We should note that another way to fix the amplitude of power spectrum is by measuring the variance of mass in a specific scale, which is conventionally introduced to be in scale of $8 Mpc/h$. In App.(\ref{App:power}) we plot the matter power spectrum for inflationary models we introduced in Sec.(\ref{Sec-1}).
To finalize this subsection, we have to study the change of the variance in matter density contrast. The variance of matter perturbations has a crucial role in the non-linear structure formation and the abundance of the structures.
The variance of mass is related to the matter power spectrum by the relation
\be \label{eq:sigma}
\sigma^2(M,z)=\int\frac{d^3k}{(2\pi)^3}P_{m}(k,z)W^2(kR),
\ee
where in this definition $P_m$  is the matter power spectrum in the linear regime introduced in Eq.(\ref{eq:matterpower}) which is obtained by using a linear theory transfer function. The window function $W(kR)$  is the Fourier transform of the top-hat filter which is defined as
\be
W(y)=3\frac{\sin(y)- y \cos(y)}{y^3},
\ee
where $y=kR$ and the mass is related to window function scale via $M=4\pi\rho R^3/3$.
\begin{figure}
\centering
\includegraphics[width=0.45\textwidth]{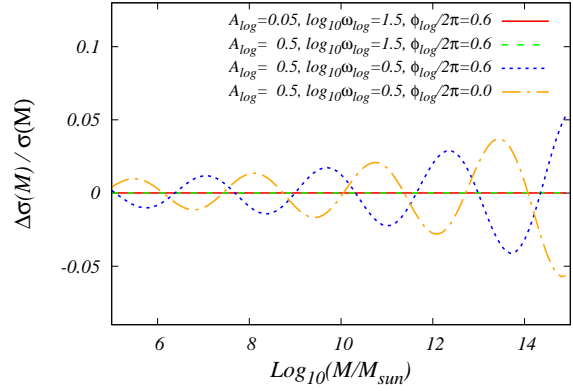}
\caption{The relative variance change is plotted versus mass for logarithmic oscillation model.  The solid red line is plotted for deviation from standard case, (log-oscillation extension) with $A_{log} = 0.03$, $log_{10}^{\omega_{log}} = 1.5$ and $\phi_{log} /2\pi = 0.6$ which are the best fit parameters with Planck data. The dashed green line is plotted with parameters $A_{log} = 0.5$, $log_{10}^{\omega_{log}} = 1.5$ and $\phi_{log} /2\pi = 0.6$. The dotted blue line is plotted with $A_{log} = 0.5$, $log_{10}^{\omega_{log}} = 0.5$ and $\phi_{log} /2\pi = 0.6$ and the orange long dash-dot line is plotted with $A_{log} = 0.5$, $log_{10}^{\omega_{log}} = 0.5$ and $\phi_{log} /2\pi = 0.0$.}
\label{fig:sigma-log-osc}
\end{figure}
In Fig.(\ref{fig:sigma-log-osc}), we plot the change in the variance of matter density perturbations versus mass for logarithmic oscillation model. A very crucial point to indicate here, is that high frequency modification has almost negligible effect on variance change (less than $0.1\%$), this is because of the cancelation due to Eq.(\ref{eq:sigma}). In contrary, the low frequency modification has significant difference. However the low frequency modifications of primordial power is already ruled out by CMB observations (see App.(\ref{App:power})). These observations may lead to crucial consequences in inflationary model building as the frequency of the oscillations  are related to the mass of fields and their couplings in inflationary era. \\
\begin{figure}
\centering
\includegraphics[width=0.45\textwidth]{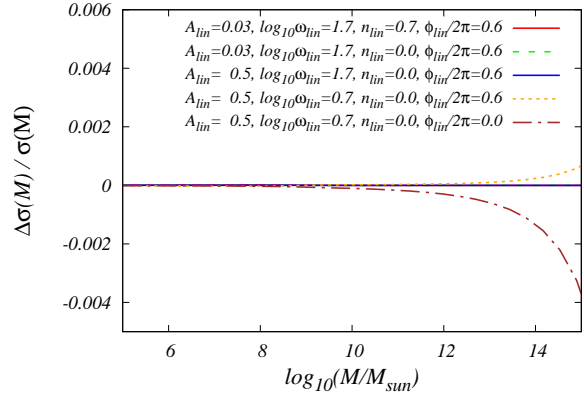}
\caption{ The relative variance change is plotted versus mass for linear oscillation model.  The solid red line is plotted for deviation from standard case, (linear-oscillation extension) with $A_{lin} = 0.03$, $log_{10}^{\omega_{lin}} = 1.7$, $n_{lin} = 0.7$ and $\phi_{log} /2\pi = 0.6$ which are the best fit parameters with Planck data.  The dashed green line is plotted with parameters $A_{lin} = 0.03$, $log_{10}^{\omega_{lin}} = 1.7$, $n_{lin} = 0.0$ and $\phi_{log} /2\pi = 0.6$. The solid blue line is plotted with $A_{lin} = 0.5$, $log_{10}^{\omega_{lin}} = 1.7$, $n_{lin} = 0.0$ and $\phi_{log} /2\pi = 0.6$. and the orange dotted line is plotted with $A_{lin} = 0.5$, $log_{10}^{\omega_{lin}} = 1.7$, $n_{lin} = 0.0$ and $\phi_{log} /2\pi = 0.6$ and finally the brown long dashed dotted line is for $A_{lin} = 0.5$, $log_{10}^{\omega_{lin}} = 1.7$, $n_{lin} = 0.0$ and $\phi_{log} /2\pi = 0.0$.}
\label{fig:sigma-lin-osc}
\end{figure}
In Fig.(\ref{fig:sigma-lin-osc}), we plot the change in the variance of matter density perturbations versus mass for linear oscillation model. In this model, again we observe that high frequency changes have negligible effect on variance term, even smaller than logarithmic oscillatory model. However the difference for low frequency is significantly higher in large masses of dark matter halos. Accordingly the effect of this model on the change of variance from the standard case can be neglected.
\begin{figure}
\centering
\includegraphics[width=0.45\textwidth]{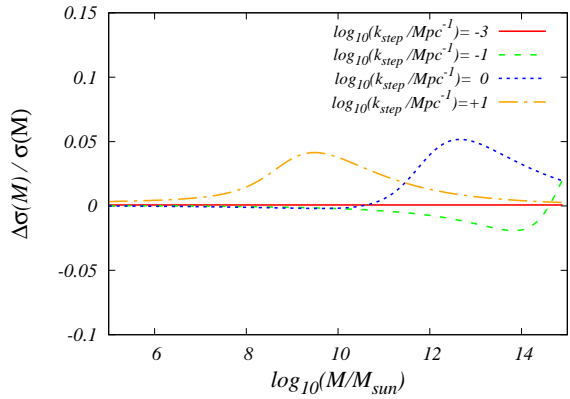}
\caption{ The ratio of variance change versus mass for inflationary models with step function. The long dashed red line is plotted for deviation from standard case, (step function) with $A_{s} = 0.374$, $\ln x_s = 0.342$ and $k_{step} = 10^{-3}/ Mpc$ which are the best fit parameters with Planck data. The long dashed green line is plotted for $k_{step} = 10^{-1}/ Mpc$, the blue dotted line is for $k_{step} = 1/Mpc$ and the orange dashed-dotted line represents $k_{step} = 10/ Mpc$ }
\label{fig:sigma-step}
\end{figure}
In Fig.(\ref{fig:sigma-step}), we plot the variance change ratio versus mass for step function in inflationary models.
This is very important result to emphasis. The specific wavenumber in which the  model breaks its scale invariance of primordial power spectrum introduce a specific mass in which the variance has the maximum change.  This specific mass scale $M_*$ can be approximated as
\be
\frac{M_*}{10^{12} M_\odot} \simeq \alpha (\frac{ Mpc /h}{ k_*})^3,
\ee
where $\alpha \simeq 1.2 \Omega_m h^2$. This mass scale corresponds to the window function radius with $R\sim1/k_*$ . This mass scale become an important indication on the number density of the structures.
In the next subsection, we will study the effect of these deviations on the number density of structures which are deduced from non-linear structure formation methods.
\subsection{Primordial Power Spectrum and Number Density of the Structures }
The mass profile of the structures can be obtained with assumption of the Press-Schechter \cite{Press:1973iz} or more sophistically by the Excursion Set Theory (EST) and the physics of random processes \cite{Bardeen:1985tr,Zentner:2006vw,Musso:2012qk, Musso:2013pha,Nikakhtar:2016bju}.
The idea of the Press-Schechter is a probabilistic approach, where it assumes that the probability of the density perturbations in initial field with a smoothing region that passes a certain barrier is almost equal to the number density of halos with the mass equal to the one in initial smoothing region.
In EST this argument is developed using the random walk mechanism of density perturbation on different scales, where it proposed a way to overcome the cloud in cloud problem. In the standard case of EST the number of trajectories that passes the barrier for the first time is related to the number density of structures in the mass scale of $M$ and $M+dM$ as
\be \label{eq:halonumberdensity}
n(M,z)=-2\frac{\rho_m}{M^2}f(\nu)\frac{d\ln\sigma(M,z)}{d\ln M},
\ee
where  $\nu \equiv \delta_c / \sigma(M,z)$ is the height parameter that is defined as the ratio of the spherical collapse critical density $\delta_c\simeq 1.68$ \cite{Gunn:1972sv} to the variance of mass $M$, which is related to the smoothing scale $R$ in window function($M=4\pi\rho R^3/3$). In the case of using sharp k-space filter for smoothing the densities , which correspond to uncorrelated steps in random walks in EST, the universality function becomes as
\be \label{PS}
f_{PS}(\nu) = \frac{1}{\sqrt{2\pi}} \nu e^{-\nu^2 / 2}.
\ee
Accordingly by knowing the primordial power spectrum and choosing the corresponding smoothing function, where in this work is sharp k-space, we can obtain the number density of structures. We should note that any deviation from standard model of collapse will introduce a moving barrier \cite{DeSimone:2010mu}. Also the change in smoothing  scale will be reflected in EST by having non-Markovian walks \cite{Musso:2014jda}. For a specific case for studying the effect of featured inflationary models, we use the  Sheth and Tormen universality function which is defined to incorporate the ellipsoidal collapse \cite{Sheth:2001dp}. The universality function of ellipsoidal collapse is approximated as
\be \label{eq:ST}
f_{ST}(\nu) = A (1 + \frac{1}{\tilde{\nu}^{2q}}) f_{PS} (\tilde{\nu}),
\ee
where $\tilde\nu = 0.84 \nu$ and $q=0.3$. The normalization $A \sim 0.322$ is set to have the integral of $f_{ST} / \nu$ over all $\nu$ s become unity.
\begin{figure}
\centering
\includegraphics[width=0.45\textwidth]{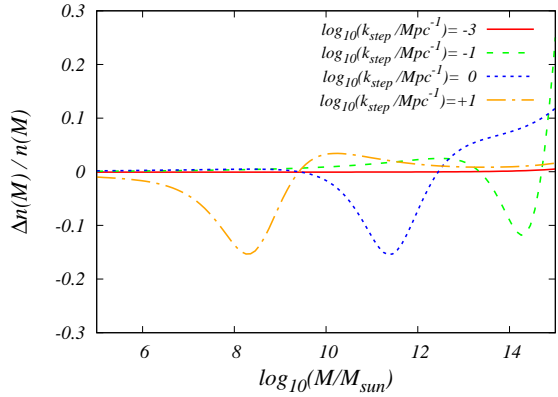}
\caption{The ratio of the difference of halo number density in comparison to $\Lambda$CDM is plotted for the step model. The red solid line is for inflationary model with step function modification in $k_{step} = 10^{-3} Mpc^{-1}$, the green long-dashed line represents the $k_{step} = 10^{-1} Mpc^{-1}$, the blue dashed line is for $k_{step} = 1 Mpc^{-1}$ and finally the $k_{step}=10 Mpc^{-1}$ is plotted with orange dashed-dotted line.}
\label{fig:nM}
\end{figure}
The number density of the structures are affected from initial power spectrum via the variance of the perturbations. In previous section, we show that oscillatory models  do not change the variance due to the integration effect over different wavenumbers. Accordingly in this subsection, we only consider the effect of the step-function model. In Fig.(\ref{fig:nM}) we plot the ratio of the change in the number density of structures in comparison with $\Lambda$CDM versus dark matter halo mass. Each curve is plotted by different position of the sharp feature. It is interesting to note that by increasing the scale of $k_{step}$, the deficit in the number of structures moves to smaller masses. As we discussed the sub-CMB scale $k>1$h/Mpc are not constrained by temperature anisotropy data, where LSS observations can be used as a test. An interesting  point to indicate in this subsection is that, the sharp feature in initial power spectrum can causes to deficit of number density of halos in specific mass range. This can be understood in a sense that the scale of the up-crossing of the trajectories is shifted. This shift is due to change in the matter power spectrum which is affected by primordial power accordingly. In the conclusion and future remarks we will discuss this point further that one of the probable solutions to the too big to fail problem \cite{BoylanKolchin:2011de,BoylanKolchin:2011dk} can be the physics of early universe which its power deviates from scale invariance in non-linear regime \cite{Kamionkowski:1999vp,Nakama:2017ohe}. However we should note that due to complicated nature of the non-linear structure formation, the effect of linear power spectrum change in the abundance of the structures is not straightforward. Accordingly N-body simulations are needed to capture the full physics of the non-linear regime with different initial conditions {\cite{Hazra:2012vs,LHuillier:2017lgm}}.
In the next subsection we will discuss the dark matter halo bias.
\subsection{Halo  Bias }
In this subsection, we discuss the effect of inflationary models on dark matter halo bias. This parameter defined as the ratio of halo number density contrast $\delta_h(M,z)$ to the underlying dark matter as
\be
b(M,z) = \delta_h(M,z) / \delta_m(z),
\ee
where $\delta_m$ is the long mode, linear regime dark matter density contrast. The linear bias in its standard definition depends on the mass of dark matter halo. In order to find this dependence we use the definition of halo density contrast $\delta_h= n(M,z; X) / \bar{n}(M,z)-1$, where $\bar{n}$ is the average number density of structures in redshift $z$ for the halo masses between $M$ and $M+dM$. The $X$ represents the parameter of the perturbation, which in the first order can be the long mode of matter perturbation. Using the Peak-background splitting idea\cite{Sheth:1999mn}, we will have
\be
b^L(M,z) = \frac{\partial \ln \bar{n} (M,z)}{\partial \delta_l},
\ee
where $b^L$ is the Lagrangian bias which is the evolved and observed to Eulerian bias, where in linear regime it becomes as $b^{E}=1+ b^{L}$. In order to obtain the bias in this framework, we use the Sheth-Tormen universality function introduced in Eq.(\ref{eq:ST}) for the number density of structures.
Again we assert that the effect of the initial power spectrum on linear regime bias is introduced via the variance, where we discussed that oscillatory models have negligible effect on it. Accordingly, we study and discuss the effect of the step-function model.
In Fig.(\ref{fig:biasstep}), we plot the difference of bias parameter from standard case with inflationary model introduced in Eq.(\ref{eq:power-step}) with the specific wavenumber $k_{step}$ in different scales. The figure shows that the bias changes in a specific mass scale depends on the scale where primordial inflationary model's potential deviates from its scale invariance potential. Also we show that when we set $k_{step} = 10^{-3}$, the change in the bias parameter is negligible, accordingly the Fisher matrix analysis presented in \cite{Chen:2016vvw} is not affected by the bias corrections.
\begin{figure}
\centering
\includegraphics[width=0.45\textwidth]{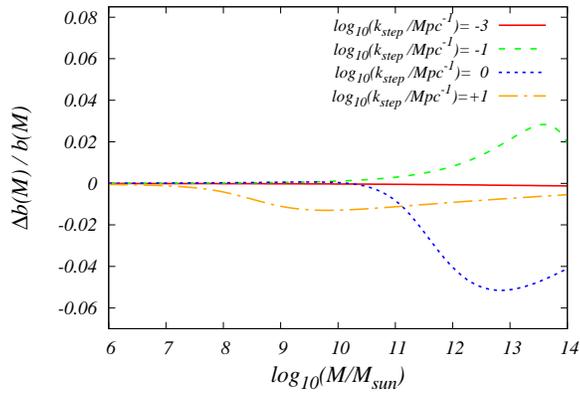}
\caption{The normalized difference of the bias from standard case is plotted versus mass. The red solid line is for inflationary model with step function modification in $k_{step} = 10^{-3} Mpc^{-1}$, the green long-dashed line represents the $k_{step} = 10^{-1} Mpc^{-1}$, the blue dashed line is for $k_{step} = 1 Mpc^{-1}$ and finally the $k_{step}=10 Mpc^{-1}$ is plotted with orange dashed-dotted line. }
\label{fig:biasstep}
\end{figure}
We should note that the bias parameter which is obtained from number density of dark matter halos due to the response effect \cite{Sheth:1999mn,Baldauf:2015vio} must be studied carefully when the effect of primordial curvature power is in small scale (non-linear regime in late time), where the complexity of non-linear structure formation can alter our conclusions.
In the next subsection, we will discuss the redshift space distortion effect, in which the bias are the main contributor for finding the correlation function and power spectrum of galaxies.
\subsection{Redshift space distortion, galaxy power spectrum and correlation function}
The velocity field of dark matter tracers, such as galaxies, by itself can be a probe of matter density field in late time sky. Due to the continuity equation in the linear regime the velocity of dark matter tracers in Fourier space is obtained as $v(k,z) = {ifH\delta_m(k,z)}/{(1+z)k}$,
where $f$ is the growth rate of the structures for late time evolution which is defined as the logarithmic derivative of density perturbation with respect to scale factor. The growth rate can be approximated as $f = \tilde{\Omega} ^\gamma(z)$ where $\tilde{\Omega} =  \Omega_m(1+z)^3 / E^2(z)$ and $\gamma \simeq 0.55$ for $\Lambda$CDM. Accordingly one can find the dispersion of the velocity field (also known as bulk flow) via the Fourier transform of the velocity power spectrum which results
\begin{equation}
\langle v^2 \rangle_R = \dfrac{H_0^2f^2}{2\pi^2}\int P(k) W^2(kR)dk.
\end{equation}
The above equation, shows how the matter density power spectrum can change the bulk flow velocity \cite{Baghram:2014qja,Habibi:2014cva}. Meanwhile the velocity field has an effect on the number density of structures in the redshift space, known as redshift space distortion effect. The galaxy power spectrum in redshift-space is as
\begin{equation}
P_g(k,z;X)= b^2(k,z;X)P_m(k,z) (1+\frac{2}{3}\beta+ \frac{1}{5}\beta^2),
\end{equation}
where $P_g(k,z;X)$ is the observed power spectrum which is calculated by a tracer $X$. The redshift space distortion parameter is $\beta = \beta(k,z;X) \equiv f / b$, where $b$ is the bias parameter. We should note that the bias depends on the characteristics of the dark matter tracer. We have to indicate that the change in the initial curvature power spectrum will change the bias parameter. Accordingly the redshift space distortion parameter $\beta$ will be altered from standard $\Lambda$CDM model. Now the interesting point to emphasis here is that the redshift space distortion parameter is used to measure the growth of structures and consequently is a test of modified gravity models. This is usually done by the combination of $f\sigma_8$ which is a bias independent quantity. However we should note that this is true in the case of $\Lambda$CDM or the models which are near to it, with almost scale independent growth function.
In Fig.(\ref{fig:biasstep}), we showed that the bias parameter can be changed due to early universe models. Accordingly the deviation from redshift-space distortion parameter or galaxy power spectrum can not be uniquely assigned to late time dark energy models \cite{Baghram:2010mc,Khosravi:2015boa}.
In consequence, the inflationary models with non standard characteristic such as sharp features can alter the redshift space distortion parameter. This degeneracy introduced in redshift-space distortion parameter by early and late time deviations of standard model.
\begin{figure}
\centering
\includegraphics[width=0.45\textwidth]{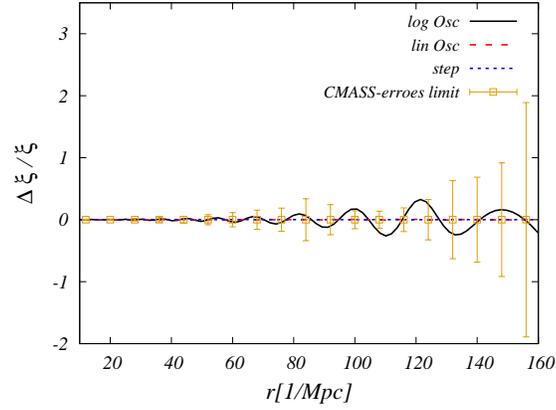}
\caption{{Difference in correlation function from $\Lambda$CDM as a function of position plotted for log oscillation (black line), linear oscillation model (red dashed line) and step function (blue double dashed line).  The inflationary models predictions is plotted by Planck best fit. The data points are from the SDSS CMASS sample, where the error-bars are plotted accordingly \cite{Gil-Marin:2015nqa}.} }
\label{Fig:correltaion}
\end{figure}
\begin{figure}
\centering
\includegraphics[width=0.45\textwidth]{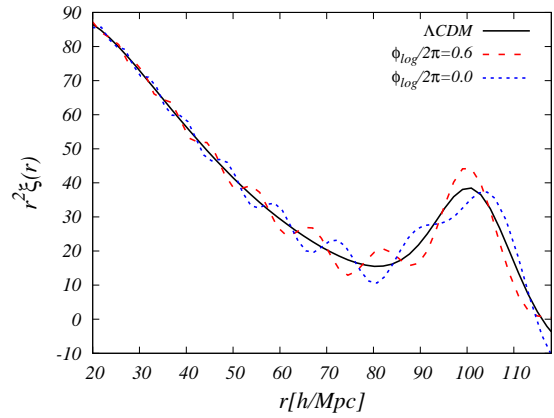}
\caption{Correlation function of matter as a function of position for $\Lambda$CDM (black solid line) and logarithmic-oscillation model
with different phases is plotted. The red long-dashed line correspond to $\phi_{log}/2\pi=0.6$, the blue dashed line is for $\phi_{log}/2\pi=0.5$ and the green dashed-dotted line represent $\phi_{log}/2\pi=0$. The amplitude and frequency of model is fixed by Planck best fit parameters.}
\label{Fig:correltaion-log}
\end{figure} \\
On the other hand one of the main observables in large scale structure, is the correlation function of dark matter tracers like galaxies and Lyman alpha forest and etc.
The interest on large scale correlation of structures becomes more because of the BAO scale which is used as a distance indicator. In this sense, we are going to study the effect of inflationary models on the correlation function and also on the BAO scale. It worths to emphasis that logarithmic and linear oscillatory models can introduce effects on BAO scale. In Fig.(\ref{Fig:correltaion}), we plot the correlation function for the standard $\Lambda$CDM model and with an inflationary model with log-oscillation, linear oscillation and step function, all of them with Planck best fit parameters. The very interesting point to indicate is that only the logarithmic-oscillation model has a drastic effect on the correlation function. This is because the model has an oscillation in primordial power and the physics of BAO oscillation imprints in transfer function where both of them affect the correlation function. Accordingly the galaxy correlation function can change the position and the amplitude of the peak. On the other hand the correlation function data like the one obtained from SDSS CMASS and LOWZ sample \cite{Gil-Marin:2015nqa} can constrain the log-oscillation models, however this analysis should be done by much care, and by including all the free parameters of the model and the reconstruction must be done carefully. The data points in Fig.(\ref{Fig:correltaion}), shows the error bars in matter correlation function.
In Fig.(\ref{Fig:correltaion-log}), we plot the correlation function restated in terms of $r^2\xi$ versus comoving distance for log-oscillatory mode with best fit parameters of Planck. The figure shows that the position and amplitude of the theoretical model of correlation function is affected by the phase of the log-oscillatory model. This change in position of the BAO scale is very important as it is used as distance indicator to distinguish between dark energy models\cite{Bassett:2009mm}.
As a final point in this subsection we should note that the non-linear effect in principle can smears out the oscillatory effect by mode mixing. This can be an obstacle for detecting the effect of oscillatory models in large scale observables, such as BAO peak\cite{LHuillier:2017lgm}.\\
In the next subsection we will discuss about the CMB-lensing as a probe of inflationary models.

\section{CMB-lensing}
\label{Sec-3}
In this section, we search for the imprints of the primordial power spectrum features on the CMB lensing map.
The weak gravitational lensing of CMB is happened due to the deflection of the CMB photons caused by the structures in the late time universe in the line of sight of the CMB photons. The interesting point about CMB lensing, (lensing in general) is that the displacement, magnification and shear of the sources is introduced by the whole matter in the line of sight ( dark matter and baryonic). This means that by using the lensing observations we can by-pass the problem of dark matter-baryonic matter bias. Accordingly the CMB lensing can be used as a fair probe of dark matter distribution which has the imprint of early universe on it. In the upcoming two subsections, we discuss the current and future status of CMB lensing observation considering the effect of the models we considered.\\
\subsection{Current Status}
The lensed  CMB temperature we observed in a specific direction is modified as $\tilde{T} (\hat{n}) = T(\hat{n}' +\vec{\alpha})$, where $\hat{n}$ indicate the direction of observation on CMB sky and $\hat{n}'$ is the direction of primordial temperature anisotropy which is mapped by the deflection angle $\vec{\alpha}$ to the observed point. The deflection angle can be extracted by $\vec{\alpha} = \vec{\nabla} \psi$, where $\psi$ is the lensing potential which can be defined for each observational direction \cite{Hassani:2015zat}
\begin{equation}
   \psi(\hat{n}) = -2 \int^{\chi_{\ast}}_0{\frac{\chi_{\ast}-\chi}{\chi_{\ast} \chi}\Phi(\chi \hat{n} ; \eta_0-\chi)}d\chi,
\end{equation}
where $\Phi$ is the Bardeen potential, $\chi_{\ast}$ is the comoving distance to the CMB, $\eta_0$ is the conformal present time($c=1$)
and the integral is over the comoving distance $\chi$.
The expansion of lensing potential in terms of spherical harmonics gives
\begin{equation}
\psi(\hat{n}) = \sum_{lm} \psi_{lm} Y_{lm} (\hat{n}),
\end{equation}
where $\psi_{lm}$ is the expansion coefficient. The physics of  the CMB lensing is encoded
in angular power spectrum which can be expressed by
\begin{equation}
\langle \psi_{lm} \psi^{\ast}_{l'm'} \rangle = \delta_{ll'} \delta{mm'} C^{\psi}_l.
\end{equation}
The lensing potential is related to the Bardeen potential and as we discussed previously
the gravitational potential is obtained from curvature perturbations by
transfer function $T(k)$ (which consider the effects of horizon crossing and matter-radiation equality) and growth function $D(z)$ (which consider the effect of late time accelerated expansion of cosmos).
On the other hand, one can express the lensing potential due to matter power spectrum. The Poisson equation relates the Bardeen potential to
matter density contrast $\Phi(k,z) = \dfrac{3}{2} \dfrac{H^2 \Omega_m(1+z)}{k^2} \delta_m$. Accordingly the lensing angular power spectrum can be written in terms of matter power spectrum as
\begin{equation} \label{eq:cmblens}
l^4 C^{\psi}_l = \int^{z_{\ast}}_0 dz W^{\psi}(z) P_{m}(\dfrac{l}{\chi},z),
\end{equation}
where $W^{\psi}$ is the kernel for lensing potential and $P_{m}$ is the
matter power spectrum, which is related to curvature power spectrum via Eq.(\ref{eq:matterpower}). Note that in obtaining Eq.(\ref{eq:cmblens}), we used the Limber approximation, which is valid for high angular moments ($l\gg1$).
Also by going to higher moments means that we are probing the matter power spectrum in non-linear regimes. Accordingly the matter power spectrum in Eq.(\ref{eq:cmblens}) must be replaced by non-linear power spectrum.
Finally we should note that the lensing kernel is
\begin{equation}
W^{\psi}(z) = 9(cH_0^{-1})^3 \dfrac{1}{E(z)} (1-\dfrac{\chi}{\chi_{\ast}})^2 \Omega_m^2(1+z)^2.
\end{equation}
\begin{figure}
\centering
\includegraphics[width=0.45\textwidth]{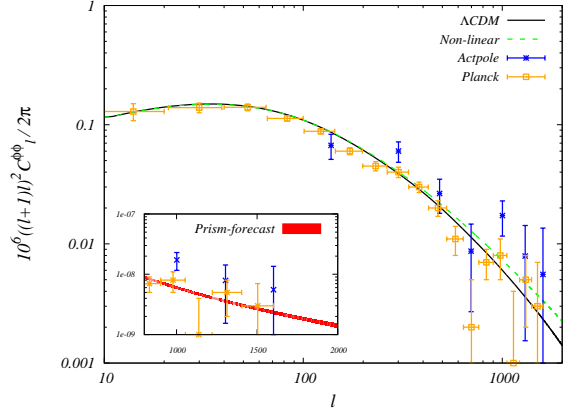}
\caption{ CMB lensing potential angular power spectrum versus moments for standard $\Lambda$CDM model with best fit parameters of Planck is plotted. The data points are from CMB lensing observation of Planck collaboration \cite{Ade:2015zua} and ACT data \cite{Sherwin:2016tyf}. The green dashed line is obtained by using the non-linear matter power spectrum \cite{Takahashi:2012em}. In the inset figure, we show the confidence level which can be achieved from future CMB experiment the PRISM collaboration.}
\label{fig:Lensing}
\end{figure}
In Fig.(\ref{fig:Lensing}), we plot the angular power spectrum of lensing potential in terms of moments $\ell$ for standard $\Lambda$CDM model with best fit parameters of Planck. The data points are from CMB lensing observation of Planck collaboration \cite{Ade:2015zua} and ACT data \cite{Sherwin:2016tyf}. The green dashed line is obtained from non-linear power spectrum \cite{Takahashi:2012em}. In the inset figure, we show the confidence level which can be achieved from future CMB experiment the PRISM collaboration\cite{Andre:2013nfa,Kasanda:2014fxa}. We should note that the non-linear effects enhance the CMB lensing effect,  this can be an interesting point for detection of the small scale features of inflationary models. However we know that the complexity of the non-linear structure formation can change the prediction that we present in this section. Complimentary N-body simulations can address this problem.
\begin{figure}
\centering
\includegraphics[width=0.45\textwidth]{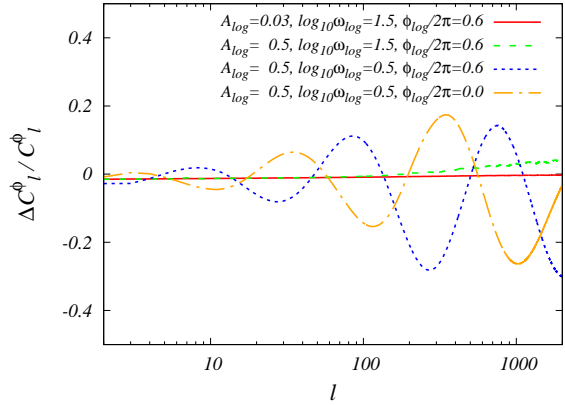}
\caption{The normalized CMB lensing potential angular power spectrum's difference from standard  case is plotted versus moments for logarithmic oscillating  inflationary models.  The parameters of logarithmic-oscillatory model for each color are the same as Fig.(\ref{fig:sigma-log-osc}).}
\label{fig:Lensing-log-Osc.eps}
\end{figure}\\
In Fig.(\ref{fig:Lensing-log-Osc.eps}), we plot the change in the angular power of CMB lensing potential with respect to the $\Lambda$CDM for logarithmic-oscillatory models with Planck best fit parameters. It is obvious that the change in lensing potential angular power spectrum with respect to Planck best fit set of parameters is negligible when the frequency of oscillations in inflationary models are high.
\begin{figure}
\centering
\includegraphics[width=0.45\textwidth]{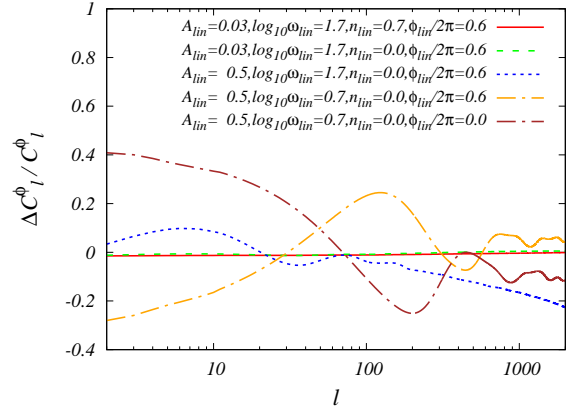}
\caption{ The normalized CMB lensing potential angular power spectrum's difference from standard case is plotted versus moments for linear oscillating  inflationary models. The parameters of logarithmic-oscillatory model for each color are the same as Fig.(\ref{fig:sigma-lin-osc}).}
\label{fig:Lensing-lin-Osc}
\end{figure}\\
In Fig.(\ref{fig:Lensing-lin-Osc}), we plot the change in angular power spectrum of CMB lensing potential with respect to $\Lambda$CDM when we choose linear-oscillatory models. The parameters are from Planck best fit and the ones which we used them in the Fig.(\ref{fig:Lensing-lin-Osc}). In this case, the same as the logarithmic-oscillatory models best fit parameters of Planck are showing non significance imprint on lensing potential angular power spectrum.
\begin{figure}
\centering
\includegraphics[width=0.45\textwidth]{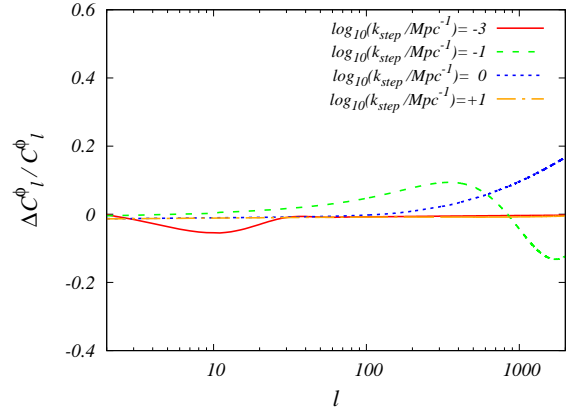}
\caption{ The normalized CMB lensing potential angular power spectrum's difference from standard  case is plotted versus moments for step function inflationary models.The parameters of logarithmic-oscillatory model for each color are the same as Fig.(\ref{fig:sigma-step}).}
\label{fig:Lensing-step}
\end{figure}
In Fig.(\ref{fig:Lensing-step}) the change in angular power spectrum of CMB lensing with respect to $\Lambda$CDM, is plotted. In which the inflationary model with step in potential is studied accordingly. In Fig.(\ref{fig:Lensing-step}) we change the $k_{s}$ which parameterize the place of the feature on primordial power spectrum. Fig.(\ref{fig:Lensing-step}) shows that deviation from $\Lambda$CDM for step in potential models is also a localized feature. The place of this localized feature depend on $k_{s}$ for high $k_{s}$ the feature on CMB lensing potential appears in high $l$ and vise versa. On the other hand, when cosmic variance is dominant in low $l$ so step in potential model with Planck best fit set of parameters, could hardly be observed and constrained with CMB lensing. Accordingly the CMB lensing is one of the best observational probes for localized feature models when the feature occurs in intermediate to large wavenumbers.

\subsection{Future Forecast}
In this subsection, we study the future forecast on the free parameters of inflationary models using the CMB -lensing observations.
As we discussed in previous sections, CMB lensing observations is one of the unique opportunities which is probing the cosmological models without the degeneracy of dark matter-baryon bias, which in its own can be effected by cosmological models and introduce some degeneracies.\\
For doing forecast we use Fisher matrix analysis. The formalism of Fisher matrix analysis of CMB lensing is introduced in App.(\ref{App:fisher matrx}). The forecast is done for future CMB experiment and also Planck experiment for comparison. In the App.(\ref{App:fisher matrx}) the result for logarithmic-oscillation, linear oscillation and step in potential model with Planck temperature map best fit parameter is calculated. As we except there is no significance result in these cases for Planck like experiments. Our result will show that in future experiments the situation is not change very much.\\
\begin{figure}
\centering
\includegraphics[width=0.45\textwidth]{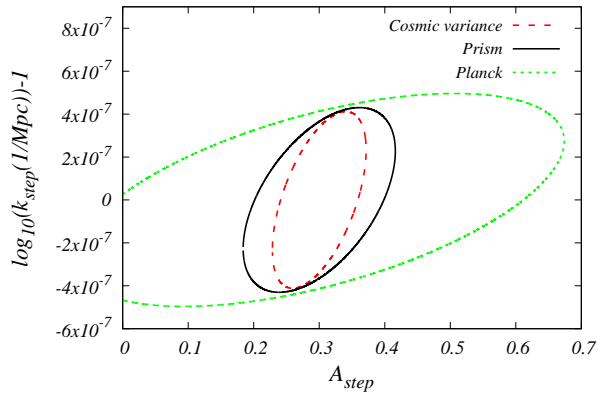}
\caption{Fisher matrix analysis of step in potential model the fiducial parameter is $k_{step} =10$ when other parameters of model is set by Planck temperature best fit.}
\label{fig:Fisher-step3}
\end{figure}
However changing the fiducial parameters may changes the situation. In Fig.(\ref{fig:Fisher-step3}) we change the $k_{step}$ from Planck best fit case to $k_{step} = 10$ when all the other parameters are unchanged. The Fig.(\ref{fig:Fisher-step3}) shows that the contours will be closed in future CMB experiment when they are open in the Planck experiment. The result is significance. It shows that a localized feature, appeared in $k \sim 10$ will be detectable in future CMB lensing experiment which is in very small scales. Deviation of primordial power spectrum from $\Lambda$CDM in these scales could be related to small scale crisis of $\Lambda$CDM.
\section{Conclusion and Future Remarks}
\label{Sec-Conc}
The CMB and LSS observations are two main arenas to test the physics of early universe. This can be done by studying  the effect of deviation from standard case of isotropic, nearly scale invariant, Gaussian and adiabatic primordial power spectrum on the observables.\\
In this work, we study the inflationary models with logarithmic and linear oscillation and a step function. We study the effect of the parameter space on CMB temperature anisotropy and late time observables. We stated and show that LSS can be used as a complimentary probe with the CMB data to study the inflationary models. Especially LSS can probe the deviations from standard initial conditions in sub-CMB scales $k > 0.3 h/Mpc$. For this statement we show that the step-like models that have a feature in small scale can have an observable effect on small scales. On the other hand we show that the oscillatory models with high frequency which is now favored by CMB observations  almost have no effect on non linear structure formation. The corresponding observations such
as number density of dark matter halos are unaffected by these models. However a very specific observation known as the correlation function of dark matter tracers can be affected from logarithmic-oscillatory models. This is a very new feature that we show in this work that the standard ruler of BAO can be affected by early universe physics. The previously known features are the dark matter-baryonic matter bias, non linearities and redshift space distortion \cite{Noh:2009bb}, where now we assert that the deviation from standard scale invariant initial conditions can have a degenerate effect on BAO scale and its amplitude. \\
Beside the BAO scale, in  this work we show that some degeneracies can be arisen in cosmological observations such as redshift space distortion and matter correlation function from the physics of early universe and the deviation from $\Lambda$CDM in late time. Accordingly we should use complementary observations to break the degeneracy. \\
On the other hand, we show that the localized features can have a significant effect on non-linear structure formation. The sub-CMB scales which are not constrained by CMB observations are open to this type of modification. More explicitly we show that the step function in $k> 1 h/Mpc$ change the number of structures. Accordingly we will propose that the galactic scale crisis of cold dark matter model can be addressed by the localized feature inflationary models with significant amplitude. This will be a proposal for future research on studying the whole physics of structure formation from halo formation, merger and evolution. \\
We should note that although the LSS open a new horizon with more data in 3D tomography of the universe for checking initial conditions, however it suffers from non-linear physics and bias of baryon-dark matter. In this direction we suggest that observations such as CMB lensing which trace the distribution of all matter is a profound one to probe the early universe model. In consequence of this we study the Fisher forecast of these models with future CMB lensing observations, which tighten the constrained on free parameter of the model. Also we show that the results strictly depend on fiducial parameters. We show that  the CMB lensing is the best experiment to detect the localized features in scales smaller than $k \sim1$ h/Mpc which are important in physics of galaxy scale. We assert that the CMB lensing is a powerful observation because it probes smaller scales in comparison to the temperature anisotropy maps. This is because the lenses spread all the way from CMB to us effectively deflect the CMB light in lower redshifts. So with the same angular resolution that we observe the CMB we can probe the smaller scales. As a specific example, we explicitly show that the future CMB observations like PRISM can detect a step-like feature in primordial power spectrum in scales of $k=10$ h/Mpc. \\
Finally we conclude that in the case that any anomaly from standard initial condition, which is found in CMB scale or it is hidden in sub-CMB scales must have its fingerprint on LSS observations, because of the whole picture of the structure formation and the future LSS surveys with small scale high resolution CMB experiments can find these effects. For future remarks, we should also note that in this work, we study the LSS probes which are related to two point function statistics. Accordingly higher order effects like non-Gaussianity worths to take into account in future studies\cite{Namjoo:2013fka,Namjoo:2014nra,Palma:2014hra,Mooij:2015cxa}
Also a full analysis of free parameters of model with the late time parameters is needed. Another direction which can be useful for further study of inflationary models can be the CMB polarization data which can help as a complimentary CMB scale probe for any deviation from standard initial conditions \cite{Namjoo:2014pqa}. The other direction which  needs a thorough investigation is the non-linear structure formation via semi-analytical methods \cite{Cooray:2002dia} or N-body simulation \cite{LHuillier:2017lgm}. The complexity of non-linear structure formation can make the detection of the features difficult, however in the case of the CMB lensing it enhances the signal. As a final word, it worths to mention again that some anomalies in late time large scale structure observations, like galactic scale crisis, can be a hint of non-standard initial conditions and can be considered as a new horizon to the physics of early universe studies.


\acknowledgments
S.B. acknowledges the hospitality of the Abdus Salam International Center for Theoretical
Physics (ICTP) during the final stage of this work.
We would like to thank  Hassan Firouzjahi, S. M. Kasanda and Ravi Sheth for their insightful comments and discussions on this work.



\begin{appendix}

\section{CMB temperature anisotropy and late time matter power spectrum}
\label{App:power}
{In this appendix, we present the study of the CMB temperature angular power spectrum for logarithmic-oscillation, linear-oscillation and step-like in potential models. The Planck data for temperature angular power spectrum and the $\Lambda$CDM best fit for matter power spectrum are also presented in figures for comparison.
\subsection{CMB temperature anisotropy}
The upcoming plots show how well the CMB constrains the deviation from standard model and in the main body of the paper, we also discuss the effect of different modification on CMB and LSS observables. }\\
\begin{figure}
\centering
\includegraphics[width=0.45\textwidth]{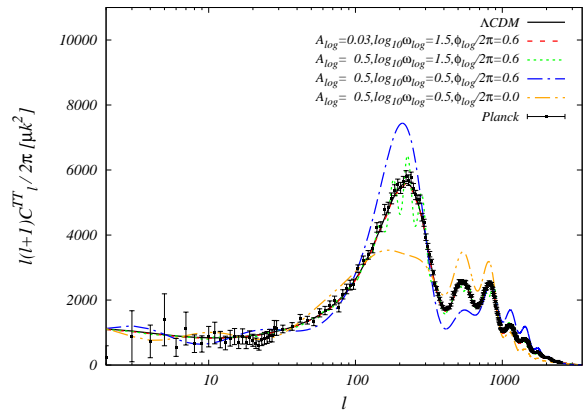}
\caption{The CMB temperature anisotropy is plotted versus moments $\ell$. The Planck data points are plotted with $1\sigma$ confidence level. The black solid line is for the $\Lambda$CDM with the standard primordial power. The parameters of logarithmic-oscillatory model for each color are the same as Fig.(\ref{fig:sigma-log-osc}) \cite{Ade:2015lrj}.}
\label{TT-log-Osc}
\end{figure}
In Fig.(\ref{TT-log-Osc}), we plot the CMB temperature anisotropy for a logarithmic-oscillation model introduced in Eq.(\ref{eq:power-log}) for the Planck best fit parameters and a set of parameters using the well known CLASS code \cite{Blas:2011rf} . The Fig.(\ref{TT-log-Osc}) shows that how strict are the constraints on this kind of models via CMB data. The main lesson we get from this figure is that the amplitude of the log-oscillation models are very constrained by CMB. This is important as we show in the main body of the work that low amplitude oscillatory models have negligible effects on the change of the number density of the structures in late time.
\begin{figure}
\centering
\includegraphics[width=0.45\textwidth]{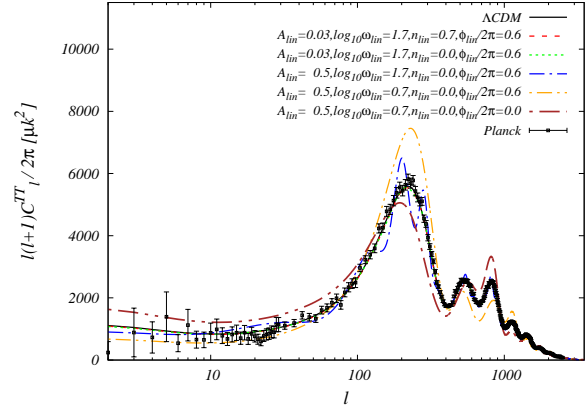}
\caption{The CMB temperature anisotropy is plotted versus moments $\ell$. The Planck data points are plotted with $1\sigma$ confidence level. The black solid line is for the $\Lambda$CDM with the standard primordial power.  The parameters of logarithmic-oscillatory model for each color are the same as Fig.(\ref{fig:sigma-lin-osc}). }
\label{Fig:Cltt-linear}
\end{figure}
In Fig.(\ref{Fig:Cltt-linear}), we plot the CMB temperature anisotropy for a linear-oscillation model introduced in Eq.(\ref{eq:power-lin}) for the Planck best fit parameters and the change of the parameters we used in this work. From this figure, again we see that the CMB temperature anisotropy constrained the deviations from standard case strongly. However in order to probe the physics of this model in late time, we will anticipate that the models could have higher amplitudes. Accordingly  We choose the spectral index of the linear oscillations correction to zero in order to not blow up the power in large scales.
In this work, we have discussed the compatibility of the models and the constrains we get from CMB and LSS, simultaneously.
\begin{figure}
\centering
\includegraphics[width=0.45\textwidth]{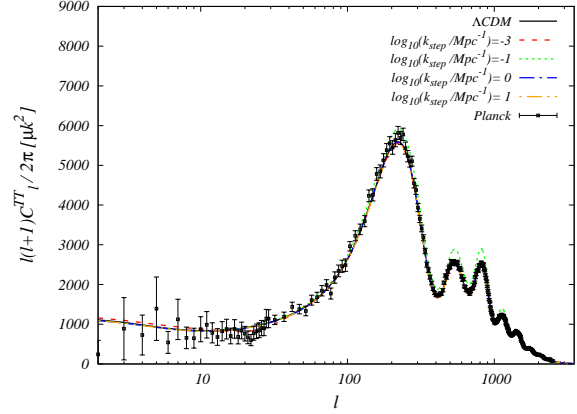}
\caption{The CMB temperature anisotropy is plotted versus moments $\ell$. The Planck data points are plotted with $1\sigma$ confidence level. The black solid line is for the $\Lambda$CDM with the standard primordial power.The parameters of logarithmic-oscillatory model for each color are the same as Fig.(\ref{fig:sigma-step}).}
\label{TT-step}
\end{figure}
In Fig.(\ref{TT-step}), we plot the CMB temperature anisotropy for a step-like  inflationary model introduced in Eq.(\ref{eq:power-step}) for the Planck best fit parameters and the change of the parameters we used in this work. The Fig. (\ref{TT-step}), shows that step in primordial power with intermediate wavenumber $k_{step} \sim {\cal{O}} (0.1) Mpc^{-1}$ are already ruled out with the temperature anisotropy data ( presented by green line in the corresponding figure).\\
However the very long wavelengths which corresponds to low $\ell$s which can address the anomalies in CMB angular power spectrum and small wavelengths which can correspond to galactic scale problems are not ruled out by CMB anisotropy data. The LSS analysis, as discussed in the main body of the work can shed light on the subject.
\subsection{Matter power spectrum}
{In this subsection we will study the imprint of inflationary model's  primordial power spectrum on the late time matter power spectrum. The matter power spectrum is calculated using linear theory and Eq.(\ref{eq:matterpower}). The Planck best fit set of parameters and the other parameter sets are considered}\\
\begin{figure}
\centering
\includegraphics[width=0.45\textwidth]{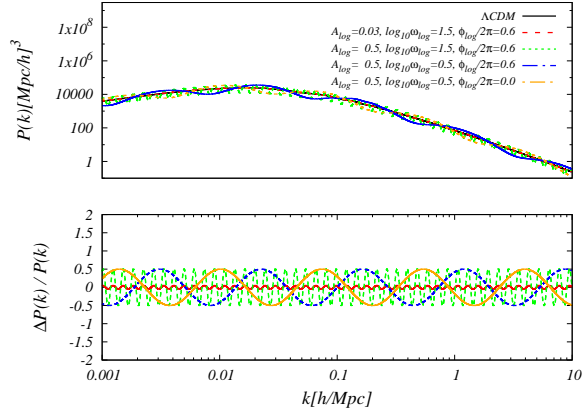}
\caption{Logarithmic oscillatory model's matter power spectrum (top) and difference with $\Lambda$CDM (bottom) with
various choice in model parameters is plotted versus wavenumber. The parameters of logarithmic-oscillatory model for each color are the same as Fig.(\ref{fig:sigma-log-osc}). }
\label{Fig:powerlog}
\end{figure}
In Fig. (\ref{Fig:powerlog}), we plot the linear matter power spectrum for $\Lambda$CDM with standard initial conditions and the power spectrum with logarithmic oscillatory primordial curvature perturbations. The plot consider the Planck best fit parameters $A_{log}\simeq 0.3$, $log_{10}\omega = 1.5$ and phase of $\phi = 2\pi\times 0.6$. The bottom panel shows the ratio of the difference in power spectrum versus wavenumber. The difference with best fit of Planck parameters on power spectrum is in order of $ \sim 1 \%$. The other curves are plotted with higher amplitude $A_{log} = 0.5$ in order to study the physics of the modification and its effect on the matter power spectrum and the other observables of large scale structure. A very important point to indicate here is that the oscillatory feature with high frequency in matter power spectrum causes to a cancelation of any desired quantity which is defined as integral over different wavenumbers, this is true for example for variance of matter perturbations. This is discussed extensively in the main body of the work.\\
\begin{figure}
\centering
\includegraphics[width=0.45\textwidth]{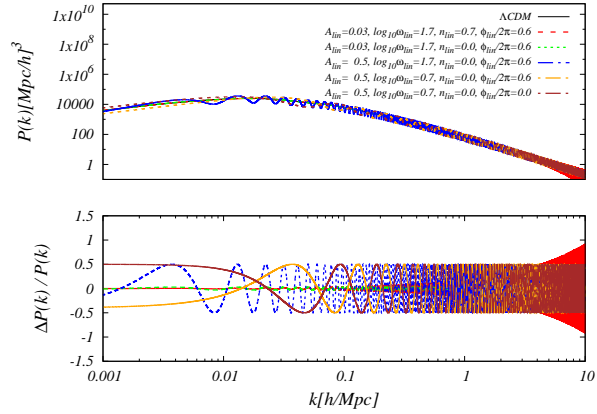}
\caption{Matter power spectrum for linear oscillatory model (top) and the difference with $\Lambda$CDM model (bottom)
with various choice in model parameters is plotted versus wavenumber. The parameters of linear-oscillatory model for each color are the same as Fig.(\ref{fig:sigma-lin-osc}). }
\label{Fig:powerlin}
\end{figure}
In Fig. (\ref{Fig:powerlin}), we plot the linear matter power spectrum for $\Lambda$CDM with standard initial conditions and the power spectrum with linear oscillatory primordial curvature perturbations. The plot consider the Planck best fit parameters $A_{lin}\simeq 0.03$, $log_{10}\omega_{lin} = 1.7$, $n_{lin} = 0.7$ and $\phi_{lin}=2\pi\times0.6$. In the linear case, also we have the same observation that the modifications with high frequency is also have no effect on integrated quantities.

\begin{figure}
\centering
\includegraphics[width=0.45\textwidth]{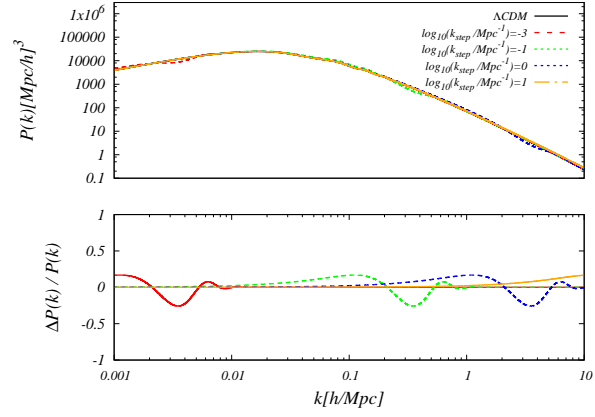}
\caption{Matter power spectrum for step function model (top) and the difference with $\Lambda$CDM model (bottom)
with various choice in model parameters is plotted versus wavenumber. The parameters of step function model for each color are the same as Fig.(\ref{fig:sigma-step}).}
\label{Fig:step}
\end{figure}
In Fig. (\ref{Fig:step}), we plot the linear matter power spectrum for the step function.
The matter power spectrum has a localized significant change due to the primordial feature. This is in contrast with the logarithmic and linear oscillation models, where they need low frequencies for their oscillations in order to be have a localized feature in late time observations.\\
This key observation will lead to a distinguishing feature which can help us to find the effect of inflationary models with step function in late time observables. Also Fig.(\ref{Fig:step}) shows that the effect of changing $k_{step}$ on power spectrum changes the position of the feature in $k$ space. \\
The main contribution of these models on late time observables are discussed extensively in Sec.(\ref{Sec-3}).

\section{Halo number density}
In this appendix we discuss some other issues on the effect of inflationary models on the number density of the structures. This is done by considering the effect of deviation from standard case in the number density in different redshifts. Also we study the effect of ellipsoidal collapse and its differences with spherical one. The halo number density for ellipsoidal collapse is obtained by using the universality function introduced in Eq.(\ref{eq:ST}) by Sheth and Tormen. This is done in order to see how ellipsoidal assumption would changes the result for halo number density. In the top panel of Fig. (\ref{fig:PS-ST}) halo number density and in the bottom panel of the figure the relative changes in halo number density with respect to $\Lambda$CDM is plotted. The model we consider here is the step function model.
We set $k_{step} =10 h/Mpc$ and the other parameters of the model is fixed by Planck best fit and both the Press-Schechter and Sheth-Tormen cases are considered. The bottom panel of Fig. (\ref{fig:PS-ST}) shows that both cases have the same features in number density due to step function.\\
\begin{figure}
\centering
\includegraphics[width=0.45\textwidth]{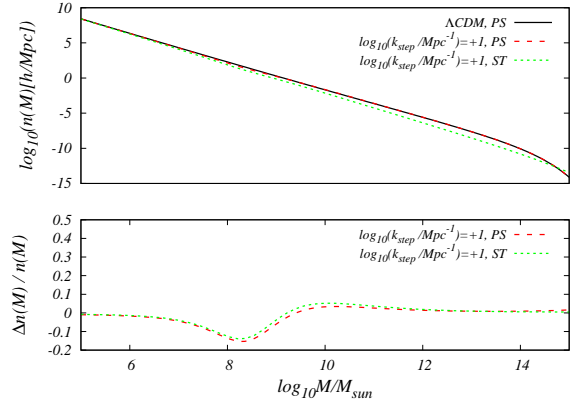}
\caption{Number density of dark matter halo for step function model and the $\Lambda$CDM and its normalized ratio is plotted versus the mass of halos. The black solid line represent the number density of structures in standard $\Lambda$CDM model case with the Press Schechter (PS) universality function. The red long dashed one represents the initial conditions with step function wavenumber $\log_{10}k_{step}/Mpc^{-1}=+1$ and with considering the PS universality function. The blue dashed line is for the same model with Sheth-Tormen (ST) function. }
\label{fig:PS-ST}
\end{figure}
{In Fig.(\ref{fig:n-step-z.eps}) we examined the halo number density (top panel) and the changes in halo number density from $\Lambda$CDM due step model (bottom panel), in different redshifts. The model for feature is the step function model in potential with $k_{step} = 10 h/Mpc$ and other parameters are set with Planck best fit. The result is calculated considering growth function in higher redshift Eq.(\ref{eq:matterpower1}) and using Eq.(\ref{eq:halonumberdensity}). The Fig.(\ref{fig:n-step-z.eps}) indicates that the deviation of number density from $\Lambda$CDM have a generic shape which is not altered strongly even in larger redshifts. }
\begin{figure}
\centering
\includegraphics[width=0.45\textwidth]{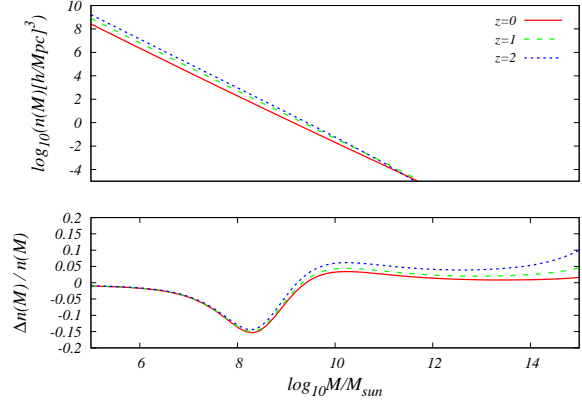}
\caption{Difference of halo number density from $\Lambda$CDM for step function model for different redshifts is plotted. The green solid line is for $z=0$. The green long dashed one represent the number density of structures in $z=1$ and $z=2$ is plotted in blue dashed line. }
\label{fig:n-step-z.eps}
\end{figure}
These generic shapes that we see in the change of the number density of structures due to features, can be a reasonable hint to the idea that we proposed in the main body of the paper that the inflationary models can address some galactic scale crisis.

\section{Fisher analysis of CMB lensing}
\label{App:fisher matrx}
\begin{figure}
\centering
\includegraphics[width=0.45\textwidth]{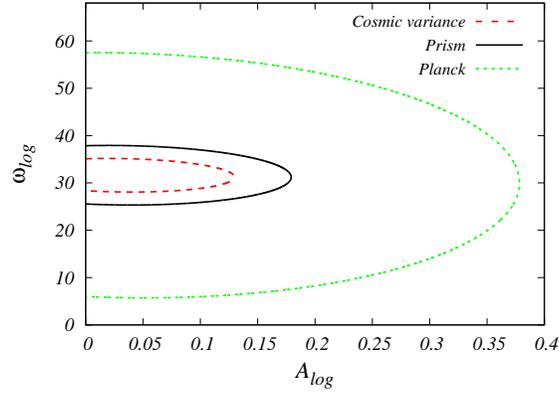}
\caption{The futuristic joint constraints on free parameters of logarithmic  oscillatory models $\log_{10}\omega_{log}$ and $A_{log}$ via CMB lensing observation is plotted. The black solid line shows the confidence level due to Prism-like observation. The green dashed line represent the Planck -like constraint and red long dashed lines shows the constraints with omitting the cosmic variance error .}
\label{fig:Fisher-log.eps}
\end{figure}
\begin{figure}[t]
\centering
\includegraphics[width=0.45\textwidth]{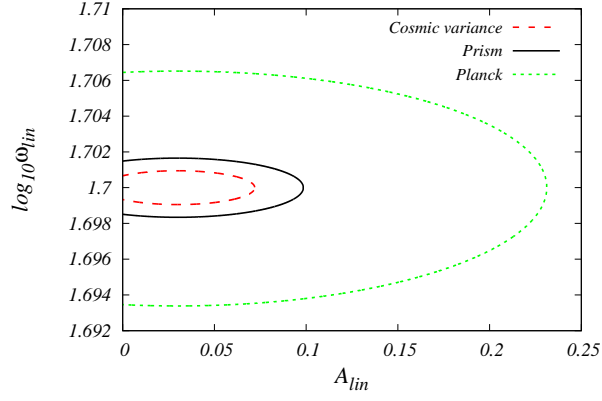}
\caption{The futuristic joint constraints on free parameters of linear oscillatory models $\log_{10}\omega_{lin}$, $A_{lin}$ via CMB lensing observation is plotted. The black solid line shows the confidence level due to Prism-like observation. The green dashed line represent the Planck -like constraint and red long dashed lines shows the constraints with omitting the cosmic variance error. }
\label{fig:Fisher-lin}
\end{figure}

In this appendix we discuss the main analysis we used for Fisher forecast of the CMB lensing probe.
Fisher matrix analysis is a simple statistical method for studying the future experiments ability to constrain models. Here we consider the special case of CMB lensing and inflationary models which are examined in this paper. The key assumption in this formalism is that the likelihood function is a multivariate Gaussian function of the parameters of the model. This proposal must be used with care for the models that their free parameters are not Gaussian around the best fit. However the Fisher matrix analysis is fast and simple method which can anticipate the result of the future experiments, when we set the parameters of models as previously known.
As mentioned, by the assumption of Fisher matrix analysis, Likelihood would be like
\begin{equation}
	L(\theta_{i}) = \exp \left[ (\theta_{i}-\bar{\theta}_{i}) F_{ij}  (\theta_j-\bar{\theta}_j) \right],
\end{equation}
where $F_{ij}$ is Fisher matrix and $\theta_i (\bar{\theta}_i)$ are parameters (fiducial parameters) of our model.
So by considering the partial derivatives of the likelihood we can derive the equation for Fisher matrix as
\begin{equation}
F_{ij} = \dfrac{\partial^2 \ln L }{\partial \theta_i \partial \theta_j}.
\end{equation}
On the other hand we can relate Likelihood to $\chi$-square of a data set. If we suppose that $f_x^o$ is a data set
for observation over different $x$ and if we calculate $f_x^o$ using our model $f^{th}_x(\theta_i)$ for a set of
parameters $\theta_i$, then by definition we can find  the Likelihood as
\begin{equation}
L(\theta_i) = e^{-\chi^2/2},
\end{equation}
where
\begin{equation}
\chi ^2 = \sum_x (f^{th}_x(\theta_i) -f^o_x) COV^{-1}_{xy} (f^{th}_y(\theta_i) -f^o_y),
\end{equation}
where $COV_{xy}$ is the correlated error matrix and is highly related to errors introduced by instruments of experiment.\\
For the CMB experiments the Fisher matrix could be find as (\cite{Eisenstein:1998hr})
\begin{equation}
F_{ij}^{XY} = \sum_l \dfrac{\partial C_l^X}{\partial \theta_i} (COV_l)^{-1}_{XY} \dfrac{\partial C_l^Y}{\partial \theta_j},
\end{equation}
here $X$ and $Y$ are introduced as temperature or $E$ mode polarization. So we neglected B-mode polarization because in usual case one could consider that B-mode polarization and lensing potential are not correlated with temperature and E-mode. So we must calculate Fisher matrix for B-mode and lensing separately. Accordingly we will have
\begin{equation}
F_{ij}^B = \sum_l \dfrac{\partial C_l^B}{\partial \theta_i} \dfrac{\partial C_l^B}{\partial \theta_j}
\dfrac{(2l+1)f_{sky}}{2(C_l^B+N_l^L)^2},
\end{equation}
and for the lensing potential spectrum we can write:
\begin{equation}
F_{ij}^L = \sum_l \dfrac{\partial C_l^L}{\partial \theta_i} \dfrac{\partial C_l^L}{\partial \theta_j}
\dfrac{(2l+1)f_{sky}}{2(C_l^L+N_l^L)^2},
\end{equation}
where in these equations $f_{sky}$ is the fraction of sky that experiment covers, the terms ${(2l+1)f_{sky}} /{2}$
is assigned for cosmic variance and $N_l$ is the detector's noise which is different for temperature and polarization and must be measured separately. For Lensing potential $N^L_l$ is the noise that arise from quadratic estimator. Finally the detector's noises are:
\begin{eqnarray} \label{eq:noise11}
N^T_l & =& \left( \dfrac{\Delta_T}{T_{CMB}} \right) e^{l(l+1) \theta_F^2/8 \ln 2},  \\ \nonumber 
N^E_l &= & N^B_l = \left( \dfrac{\Delta_P}{T_{CMB}} \right) e^{l(l+1) \theta_F^2/8 \ln 2}.
\end{eqnarray}
In which we neglect effects of foreground noises when they can be eliminated by several frequency bands measurement. In above equations $T_{CMB}$ is the CMB temperature which is set to $2.73 K$ and $\Delta_T$ and $\theta_F$ are some parameters that define the strength of the instrument. These two parameters and also the frequency band are reported as details of instruments. For Planck and future like experiment we follow Kasanda and Moodley  \cite{Kasanda:2014fxa}. \\
For lensing potential the situation is different and the noise is mainly arises from reconstruction of lensing map
from lensed temperature and polarization data. For instance the $N^L_l$ is taken to be the noise of minimum variance
estimator.\cite{Kasanda:2014fxa,Hu:2001kj}
\begin{equation}
N^L_l = \dfrac{1}{\sum_{\alpha \beta} N^{-1}_{\alpha \beta}(l) }.
\end{equation}
Note that $\alpha$ and $\beta$ represent pair of $(x,x')$ which $x$ and $x'$ are belong to $\{ T,E,B \}$. Accordingly for $N_{\alpha \beta}$ we have $6\times6=36$ components. (If we neglect BB-map the number of components will reduce to $25$). The $N_{\alpha\beta}$ is given by ~\cite{Kasanda:2014fxa,Hu:2001kj}
\begin{eqnarray}
N_{\alpha \beta} (l) &=& l^2 A_{\alpha} (l) A_{\beta}(l) \int \dfrac{d^2l_1}{(2\pi)^2} F_{\alpha}(l_1,l_2) \\ \nonumber
&[& F_{\beta}(l_1,l_2) \times C^{x_{\alpha} x_{\beta}}_{l_1} C^{x'_{\alpha} x'_{\beta}}_{l_2}  \\ \nonumber
&+& F_{\beta}(l_2,l_1) \times C^{x_{\alpha} x'_{\beta}}_{l_1} C^{x'_{\alpha} x_{\beta}}_{l_2} ],
\end{eqnarray}
where $l_2 = l-l_1$ and $A_{\alpha}$ and $F_{\alpha}$ are some functions that depend to angular power spectrums of $\alpha$ and $l$. As mentioned $\alpha$ (or $\beta$) are any pair of $(x,x')$ which $x$ and $x'$ are members of $\{ T,E,B \}$  \\
Also we have~\cite{Kasanda:2014fxa,Hu:2001kj}
\begin{equation}
N_{\alpha \alpha}(l) = A_{\alpha}(l).
\end{equation}
Here for simplicity we assume only TT-map is used for reconstruction and will neglect all other terms which can become correlated with polarization maps. This is useful when only temperature map is used to reconstruct lensing potential. By this simplification the noise will be as\cite{Hu:2001tn}
\begin{equation}
N^{-1}_l = \dfrac{1}{l^2} \int \dfrac{d^2l_1}{(2\pi)^2}
\dfrac{(l.l_1 \tilde{C}_{l_1} + l.l_2 \tilde{C}_{l_2})^2}{2C^{tot}_{l_1}C^{tot}_{l_2}},
\end{equation}
where $C^{tot}_l = C_l +C^{noise}_l$ and $C_l (\tilde{C}_l)$ is lensed (unlensed) power spectrum and $C_l^{noise}$ is the noise of power spectrum which is defined in Eq.(\ref{eq:noise11}). For our work $C_l = \bar{C}_L + \delta C_l$ in which $\bar{C}_l$ is related to $\Lambda$CDM and $\delta C_L$ is related to features on power spectrum.
In this work in calculation of noise, we neglect the $\delta C_l$ term and only we use the background power spectrum. This is because the main effect in Eq.(\ref{eq:noise11}) is due to $\Lambda$CDM.\\
In Fig. (\ref{fig:Fisher-log.eps}), we plot the futuristic joint confidence level on free parameters of logarithmic-oscillatory models $A_{log}$ and $\log_{10}\omega_{log}$ with the fiducial parameters of Planck best fit. For this analysis we use the configuration of future planned PRISM-like CMB observation and we compare it with a Planck-like probe. It seems that the CMB lensing in its futuristic case, as a single observation, can not distinguish the logarithmic-oscillatory models from the standard case. \\
In Fig.(\ref{fig:Fisher-lin}), we plot the futuristic joint confidence level on free parameters of linear-oscillatory models $A_{lin}$ and $\log_{10}\omega_{lin}$ with the fiducial parameters of Planck best fit. For this analysis we use the configuration of future planned PRISM-like CMB observation and we compare it with a Planck-like probe.
In Fig.(\ref{fig:Fisher-step}), we plot the constraints on the futuristic case of Prism-like observations on free parameters of the model. As the CMB lensing probe the intermediate wavenumbers, the fiducial parameters obtained from Planck observation which is $\log_{10}k_{s} / Mpc^{-1} = -3.1$ , $A_s = 0.374$ is out of the reach of CMB lensing. This is mainly because of the cosmic variance error.
\begin{figure}
\centering
\includegraphics[width=0.45\textwidth]{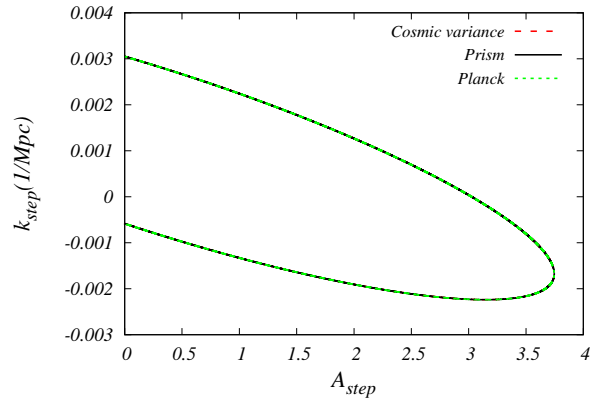}
\caption{The futuristic joint constraints on free parameters of step function models ($k_{step}$ and $A_{step}$ )via CMB lensing observation is plotted. The black solid line shows the confidence level due to Prism-like observation. The green dashed line represent the Planck -like constraint and red long dashed line shows the constraints with omitting the cosmic variance error.}
\label{fig:Fisher-step}
\end{figure}
\end{appendix}
However the very interesting case of CMB lensing with a step-like model with a feature in small scales is discussed in the main body of the paper and in Fig.(\ref{fig:Fisher-step3}).

\end{document}